\documentclass[11pt,a4paper,showpacs,showkeys,,superscriptaddress]{article}

\usepackage{epsfig}
\usepackage{amsmath,amssymb}
\usepackage{graphicx}
\usepackage{xcolor}
\usepackage{subfigure}

\usepackage{latexsym}
\usepackage{hyperref}
\hypersetup{colorlinks,%
  citecolor=blue,%
  linkcolor=red,%
  pdftex}

\begin{document}

\baselineskip 6mm
\renewcommand{\thefootnote}{\fnsymbol{footnote}}

\newcommand{\nc}{\newcommand}
\newcommand{\rnc}{\renewcommand}



\newcommand{\tcb}{\textcolor{blue}}
\newcommand{\tcr}{\textcolor{red}}
\newcommand{\tcg}{\textcolor{green}}


\def\beq{\begin{equation}}
\def\eeq{\end{equation}}
\def\ba{\begin{array}}
\def\ea{\end{array}}
\def\bea{\begin{eqnarray}}
\def\eea{\end{eqnarray}}
\def\nn{\nonumber}


\def\CMP{Commun. Math. Phys.~}
\def\JHEP{JHEP~}
\def\Pre{Preprint}
\def\PRL{Phys. Rev. Lett.~}
\def\PR {Phys. Rev.~}
\def\CQG {Class. Quant. Grav.~}
\def\PL {Phys. Lett.~}
\def\NP {Nucl. Phys.~}

\def\G{\Gamma}

\def\S{{\bf S}}
\def\C{{\bf C}}
\def\Z{{\bf Z}}
\def\R{{\bf R}}
\def\N{{\bf N}}
\def\M{{\bf M}}
\def\P{{\bf P}}
\def\bm{{\bf m}}
\def\bn{{\bf n}}

\def\CA{{\cal A}}
\def\CB{{\cal B}}
\def\CC{{\cal C}}
\def\CD{{\cal D}}
\def\CE{{\cal E}}
\def\CF{{\cal F}}
\def\CH{{\cal H}}
\def\CM{{\cal M}}
\def\CG{{\cal G}}
\def\CI{{\cal I}}
\def\CJ{{\cal J}}
\def\CL{{\cal L}}
\def\CK{{\cal K}}
\def\CN{{\cal N}}
\def\CO{{\cal O}}
\def\CP{{\cal P}}
\def\CQ{{\cal Q}}
\def\CR{{\cal R}}
\def\CS{{\cal S}}
\def\CT{{\cal T}}
\def\CU{{\cal U}}
\def\CV{{\cal V}}
\def\CW{{\cal W}}
\def\CX{{\cal X}}
\def\CY{{\cal Y}}
\def\CZ{{\cal Z}}

\def\We{{W_{\mbox{eff}}}}


\newcommand{\Lie}{\pounds}

\newcommand{\p}{\partial}
\newcommand{\bp}{\bar{\partial}}

\newcommand{\half}{\frac{1}{2}}

\newcommand{\bfalpha}{{\mbox{\boldmath $\alpha$}}}
\newcommand{\bfbeta}{{\mbox{\boldmath $\beta$}}}
\newcommand{\bfgamma}{{\mbox{\boldmath $\gamma$}}}
\newcommand{\bfmu}{{\mbox{\boldmath $\mu$}}}
\newcommand{\bfpi}{{\mbox{\boldmath $\pi$}}}
\newcommand{\bfvarpi}{{\mbox{\boldmath $\varpi$}}}
\newcommand{\bftau}{{\mbox{\boldmath $\tau$}}}
\newcommand{\bfeta}{{\mbox{\boldmath $\eta$}}}
\newcommand{\bfxi}{{\mbox{\boldmath $\xi$}}}
\newcommand{\bfkappa}{{\mbox{\boldmath $\kappa$}}}
\newcommand{\bfepsilon}{{\mbox{\boldmath $\epsilon$}}}
\newcommand{\bfTheta}{{\mbox{\boldmath $\Theta$}}}

\newcommand{\bz}{{\bar{z}}}

\newcommand{\dalpha}{\dot{\alpha}}
\newcommand{\dbeta}{\dot{\beta}}
\newcommand{\blambda}{\bar{\lambda}}
\newcommand{\btheta}{{\bar{\theta}}}
\newcommand{\bsigma}{{{\bar{\sigma}}}}
\newcommand{\bepsilon}{{\bar{\epsilon}}}
\newcommand{\bpsi}{{\bar{\psi}}}


\def\ct{\cite}
\def\la{\label}
\def\eq#1{(\ref{#1})}


\def\a{\alpha}
\def\b{\beta}
\def\g{\gamma}
\def\G{\Gamma}
\def\d{\delta}
\def\D{\Delta}
\def\ep{\epsilon}
\def\e{\eta}
\def\ph{\phi}
\def\Ph{\Phi}
\def\ps{\psi}
\def\Ps{\Psi}
\def\k{\kappa}
\def\l{\lambda}
\def\L{\Lambda}
\def\m{\mu}
\def\n{\nu}
\def\th{\theta}
\def\Th{\Theta}
\def\r{\rho}
\def\s{\sigma}
\def\S{\Sigma}
\def\ta{\tau}
\def\o{\omega}
\def\O{\Omega}
\def\pr{\prime}


\def\half{\frac{1}{2}}

\def\goto{\rightarrow}

\def\na{\nabla}
\def\grad{\nabla}
\def\curl{\nabla\times}
\def\div{\nabla\cdot}
\def\pa{\partial}

\def\bra{\left\langle}
\def\ket{\right\rangle}
\def\lb{\left[}
\def\lc{\left\{}
\def\ls{\left(}
\def\lp{\left.}
\def\rp{\right.}
\def\rb{\right]}
\def\rc{\right\}}
\def\rs{\right)}
\def\cl{\mathcal{l}}

\def\vac#1{\mid #1 \rangle}

\def\td#1{\tilde{#1}}
\def\check{ \maltese {\bf Check!}}


\def\Tr{{\rm Tr}\,}
\def\det{{\rm det}\,}


\def\bc#1{\nnindent {\bf $\bullet$ #1} \\ }
\def\ch {$<Check!>$ }
\def\ss {\vspace{1.5cm}}

\begin{titlepage}

\hfill\parbox{5cm} { }

\hskip1cm

\vspace{10mm}

\begin{center}
{\Large \bf Quasi-local conserved  charges  and holography}

\vskip 1. cm
  { Seungjoon Hyun\footnote{e-mail : sjhyun@yonsei.ac.kr}, Jaehoon Jeong\footnote{e-mail : j.jeong@yonsei.ac.kr}, Sang-A Park\footnote{e-mail : sangapark@yonsei.ac.kr},
  Sang-Heon Yi\footnote{e-mail : shyi@yonsei.ac.kr} 
  }

\vskip 0.5cm

{\it Department of Physics, College of Science, Yonsei University, Seoul 120-749, Korea}
\end{center}

\thispagestyle{empty}

\vskip1.5cm

 
\centerline{\bf ABSTRACT} \vskip 4mm
We construct  a quasi-local formalism for conserved charges in a theory of gravity  in the presence of matter fields which may have slow falloff behaviors at the asymptotic infinity. This construction depends only on equations of motion  and so it is irrespective of ambiguities  in the total derivatives of the  Lagrangian. By using identically conserved currents, we show that this formalism  leads to the same expressions of conserved charges as those in  the covariant phase space approach.  At the boundary of the asymptotic AdS space, we  also introduce an  identically conserved boundary current which has the same structure as the bulk current and then show that this boundary current gives us the holographic conserved charges identical with those from the boundary stress tensor method. 
In our quasi-local formalism  we present a general proof that conserved charges from the bulk potential  are identical with those from the boundary current. Our results can be regarded as the extension of the existing  results on the equivalence of conserved charges by the covariant phase  space approach and by the boundary stress tensor  method. \vspace{1cm} 
\noindent 
\vspace{2cm}


\end{titlepage}

\renewcommand{\thefootnote}{\arabic{footnote}}
\setcounter{footnote}{0}

\section{Introduction}

 The AdS/CFT correspondence has made huge impact on our understanding of strong coupling physics which is far beyond our usual perturbative approach in field theories.  Through this correspondence, the strongly-coupled highly-quantal regime in the dual field theory side is explored by a  classical gravity computation. Many results have been obtained in this route and various cross-checks have been made for such results verifying the power of the AdS/CFT correspondence. Its successful realization in four-dimensional supersymmetric Yang-Mills theory is still an on-going productive subject. On the other hand, the quantum gravity is not yet fully understood even under this correspondence though it may  in the future, turn out to be a crucial corner stone of the complete understanding of quantum gravity. However, the lack of its usefulness in the full quantum regime of gravity does not mean that it is powerless  in the classical theory of gravity. One of such application of the AdS/CFT correspondence to the classical gravity side is the new understanding on conserved charges in a theory of gravity.

In a theory of gravity with diffeomorphism symmetry, it is not so straightforward to define conserved charges. As is well-known, the Noether method is insufficient to connect conserved charges and symmetries when those under consideration are local gauge symmetries like diffeomorphisms. There have been various attempts  to define conserved charges in gravity and the final form of such attempts  for the asymptotically flat geometry is molded as the so-called ADM formula~\cite{Arnowitt:1962hi,Regge:1974zd}, which computes  total conserved charges at the asymptotic infinity. After failure of many attempts to construct local conserved quantities in gravity, it has been gradually recognized that  local conservation concept   like conserved currents  has intrinsic ambiguities and denies its complete specification. At most, one may try to construct quasi-local quantities in such a theory. See \cite{Szabados:2009eka} for a review on general quasi-local concepts.  We use the definition of the term {\it quasi-local} conserved charge associated with an exact Killing vector  as a surface integral in the bulk, not only at the asymptotic boundary, following the spirit given in~\cite{Wald:1993nt,Iyer:1994ys}.  One of the important results by the quasi-local construction of  conservation law is the understanding of the black hole entropy as a conserved quantity at the Killing horizon~\cite{Wald:1993nt}, which was at first perceived at the level of the analogy with thermodynamics~\cite{Bekenstein:1973ur} and then confirmed by a semi-classical computation~\cite{Hawking:1974sw}.  

In contrast to gravity, conserved charges in the dual field theory are rather clear to define and have no ambiguities in their construction. The AdS/CFT correspondence  implies that there may be a way to construct a quasi-local conserved charges in the bulk gravity side for the asymptotically AdS space, consistently with the unambiguous field theory side. Indeed, there is a formalism known as the counter term method or the boundary stress tensor method~\cite{Balasubramanian:1999re}  to obtain holographic conserved charges consistent with the dual field theory. Then, one may ask what is the relation between this holographic approach and the traditional approaches to conserved charges in gravity. This question was answered quite concretely for the asymptotically AdS geometry in Einstein gravity~\cite{Papadimitriou:2005ii,Hollands:2005wt,Hollands:2005ya}.
However, the status of this equivalence  at the general setup is not so explicit since the boundary stress tensor method depends on the explicit form of Gibbons-Hawking(GH) terms~\cite{York:1972sj,Gibbons:1976ue} and counter terms  which are not known in general. The boundary stress tensor method is  based on the  Ref.~\cite{Brown:1992br} and  is basically a kind of the Hamiltonian approach to conserved charges. Because of this nature,  this method becomes complicated for a higher derivative theory of gravity. On the other hand, in the bulk gravity side, there are general covariant methods to obtain conserved charges. For instance, the covariant phase space method~\cite{Lee:1990nz,Wald:1993nt,Iyer:1994ys,Wald:1999wa} or Barnich-Brandt-Comp\`{e}re formalism~\cite{Barnich:2001jy, Barnich:2007bf,Barnich:2003xg, Compere:2007az} can be used in a general covariant theory of gravity.  Though there is also a general argument on the consistency of the boundary stress tensor method for the asymptotically AdS geometry with the covariant phase space method~\cite{Hollands:2005ya}, it would be much better to have an explicit verification of the equivalence between the conserved charges in the holographic method and those in the bulk covariant one.

In order to verify  the equivalence of boundary and bulk formalisms for the asymptotically AdS geometry, it is useful to recall that there are some modifications on the boundary terms in the Lagrangian in the holographic method, which does not change the bulk equations of motion(EOM). Based on this fact, it is more natural to resort to a covariant formalism for conserved charges which uses the bulk EOM or more accurately the Euler-Lagrange expression. There is one such formalism developed by Abbott-Deser-Tekin(ADT)~\cite{Abbott:1981ff,Abbott:1982jh,Deser:2002rt,Deser:2002jk}, which has been used successfully for the asymptotic AdS space. 

In this paper we construct a quasi-local formalism for conserved charges in the presence of arbitrary matter fields in the theory of gravity with diffeomorphism symmetry. This construction is based on the EOM and free from any ambiguity in the total derivatives of the Lagrangian, which may be thought as the extension of the well-known ADT formalism for conserved charges. When the falloff of matter fields is slow enough, the original ADT method  needs to be extended since their approach is based on the assumption of the fast falloff of matter fields at the asymptotic infinity so that only metric contribution survives. Here we give the natural extension of the ADT formalism in the case of the slow falloff of matter fields through the construction of identically conserved currents. It turns out that this quasi-local formalism gives conserved charges which are identical with those from the covariant phase space method. Furthermore, we propose new holographic method for asymptotically AdS geometry to find the conserved charges at the boundary in the same spirit with the bulk quasi-local formalism. We show that this method gives consistent results with the boundary stress tensor method for holographic conserved charges in Einstein gravity. By using our holographic construction, we confirm the equivalence between  conserved charges in the holographic method and those in the bulk covariant one.  

This paper is organized as follows. In section 2, we construct a quasi-local formalism for  conserved charges, based on the Euler-Lagrange expressions, in the presence of arbitrary matter fields, which may be thought as the extension of the ADT formalism. We introduce the off-shell ADT current and potential and show that the resultant conserved charges are identical with those from the covariant phase space method. In section 3, we introduce the identically conserved current at the boundary and show that the corresponding conserved charges are equivalent to  those in boundary stress tensor formalism. We also
show that the boundary current  is equivalent to the bulk ADT potential in appropriate coordinates. These results warrant explicitly the equivalence of the bulk conserved charges with the holographic ones. In section 4, we summarize some generic features for scalar fields. In section 5, we apply our formalism to some interesting examples and explain additional interesting features in our formalism. In the final section, we summarize our results and comment on some future directions.

\section{Quasi-local formalism and covariant phase space approach}

In this section we extend a quasi-local formalism for  conserved charges to a theory of gravity with arbitrary   matter fields. We construct the off-shell ADT current and potential and  show that the resultant on-shell potential becomes identical with the one  from the covariant phase space method. By using these  results, we derive the ADT potential straightforwardly for a class of model, which can be used to compute conserved charges.

\subsection{Construction}

Let us consider a generic theory of gravity in the presence of  arbitrary matter fields denoted collectively as   $\psi =(\phi^I, A_{\mu}, \cdots)$,
\begin{equation} \label{}
I[g,\psi] = \frac{1}{16 \pi G}\int d^Dx \sqrt{-g}\, \CL(g,\psi)\,.
\end{equation}
For our convenience, we also denote the metric and matter fields jointly as $\Psi = (g_{\mu\nu}, \phi^{I}, A_{\mu},\cdots)$ in the following. The variation of the action would be taken as
\begin{equation} \label{}
\delta I[\Psi] = \frac{1}{16 \pi G}\int d^Dx \Big[ \sqrt{-g}\CE_{\Psi}\delta \Psi + \p_{\mu}\Theta^{\mu}(\delta \Psi)\Big]\,,
\end{equation}
where $\CE_{\Psi} =  (\CE_{\mu\nu}, \CE_{\psi})$ and $\Theta^{\mu}$ denote the Euler-Lagrange expression  and  the surface term, respectively.  We have also adopted the convention  such that  $\CE_{\Psi} \delta \Psi \equiv \CE_{\mu\nu}\delta g^{\mu\nu} + \CE_{\psi} \delta \psi$. 

In order to introduce the off-shell ADT current and potential in this generic case, we would like to note that there is an off-shell identity in the form of 
\begin{equation} \label{ida}
2\zeta_{\nu}\nabla_{\mu}\CE^{\mu\nu}+\CE_{\psi}\,  \Lie_{\zeta}\psi = \nabla_{\mu}(\CZ^{\mu\nu}\zeta_{\nu})\,,
\end{equation}
where $\Lie_{\zeta}\psi$ denotes the Lie derivative of $\psi$ along $\zeta$ direction. The second rank tensor $\CZ^{\mu\nu}$ is a certain function of metric and matter fields whose specific form will be discussed below.  This identity may be thought of as the generalization of the usual Bianchi identity. In fact, one can see that $\CZ^{\mu\nu}$ tensor vanishes when the matter EOM are satisfied by comparing the terms proportional to $\nabla_{\mu}\zeta_{\nu}$ in the left and right hand sides of Eq.~(\ref{ida}). In other words, $\CZ^{\mu\nu}$ tensor should be proportional to a certain combination of the Euler-Lagrange expression, $\CE_{\psi}$ of matter fields.  The above off-shell identity can be written in the form of 
\begin{equation} \label{idaaa}
\nabla_{\mu}\Big(2{\bf E}^{\mu\nu}\zeta_{\nu}\Big) = \CE^{\mu\nu}\Lie_{\zeta}g_{\mu\nu} - \CE_{\psi}\Lie_{\zeta}\psi=-\CE_{\Psi}\Lie_\zeta \Psi\,,
\end{equation}
where ${\bf E}^{\mu\nu}$ is defined by
\begin{equation} \label{}
{\bf E}^{\mu\nu} \equiv \CE^{\mu\nu} - \frac{1}{2}\CZ^{\mu\nu}\,.
\end{equation}
Note that the current $S^{\mu}_{\zeta} \equiv 2{\bf E}^{\mu\nu}\zeta_{\nu}$, may be identified with the weakly vanishing Noether current in Ref.~\cite{Barnich:2001jy,Barnich:2007bf,Compere:2007az}.

Some comments are in order.  
\begin{itemize}
\item The explicit form of $\CZ^{\mu\nu}$ tensor may be written as 
\[   
\sqrt{-g} \CZ^{\mu\nu}\zeta_{\nu} = \zeta^{\mu}\sqrt{-g}\CL +\Sigma^{\mu}(\zeta)  -  \Theta^{\mu}(\Lie_{\zeta} \Psi) + 2\sqrt{-g}\CE^{\mu\nu}\zeta_{\nu} + \p_{\nu}U^{\mu\nu}\,,
\]
where $U^{\mu\nu}=U^{[\mu\nu]}$ is an arbitrary antisymmetric second rank tensor. However, there are various ambiguities in this expression. We have bypassed these ambiguities by choosing $\CZ^{\m\n}$ tensor such that it is proportional to a certain combination of Euler-Lagrange expressions for matter fields, $\CE_{\psi}$. 
\item  In some cases, $\CZ^{\m\n}$ tensor turns out to vanish. Let us consider scalar fields specifically.  Since the Lie derivative of scalar fields does not contain a derivative of diffeomorphism parameter as $\Lie_{\zeta}\phi^I = \zeta^{\mu}\p_{\mu}\phi^I$, one cannot obtain terms matching with $\CZ^{\mu\nu}\nabla_{\mu}\zeta_{\nu}$ as can be inferred from Eq.~(\ref{ida}).  Therefore, $\CZ^{\mu\nu}$ tensor should vanish generically for scalar fields, though the contribution of the scalar field to $\CZ^{\mu\nu}$ may exist indirectly through the interaction with other matter fields.  In the case of massless gauge fields, if we use the modified  Lie derivative $\Lie'_{\zeta}$, supplemented by the gauge transformation,  one can see that  $\CZ^{\m\n}$ tensor vanishes. The details will be given in the following.
\item In most of interesting cases, the Lagrangian could be separated as $\CL= \CL_g + \CL_{\psi}$ for the pure gravity part  and the matter field part, respectively. 
The equations of motion of the metric and matter fields  are given by
\begin{equation} \label{}
\CE_{\mu\nu} = \CG_{\mu\nu} - T_{\mu\nu} =0\,, \qquad \CE_{\psi} =0\,,
\end{equation}
where $\CG_{\mu\nu}$ and $T^{\mu\nu}$ denote the generalized Einstein tensor  and  the stress tensor of matter fields, respectively.  
In these cases, the generalized Einstein tensor $\CG_{\mu\nu}$ for the metric field satisfies the Bianchi identity $\nabla_{\mu}\CG^{\mu\nu}=0$ and the Euler-Lagrange expression of matter fields, $\CE_{\psi}$  satisfies the following off-shell identity, independently 
\begin{equation} \label{Bulkid}
-2\zeta_{\nu}\nabla_{\mu}T^{\mu\nu}+\CE_{\psi}\,  \Lie_{\zeta}\psi = \nabla_{\mu}(\CZ^{\mu\nu}\zeta_{\nu})\,.
\end{equation}

\end{itemize}

Now,   let us recall the form of the off-shell ADT current  for a Killing vector $\xi$ in the case of pure gravity~\cite{Kim:2013zha}(See also~\cite{Bouchareb:2007yx})
\[   
J^{\mu}_{ADT} (\xi, \delta g) = \delta \CG^{\mu\nu}\xi_{\nu} + \frac{1}{2}g^{\alpha\beta}\delta g_{\alpha\beta}\, \CG^{\mu\nu}\xi_{\nu}  + \CG^{\mu\nu}\delta g_{\nu\rho}\, \xi^{\rho} + \frac{1}{2}\xi^{\mu}\CG_{\mu\nu}\delta g^{\mu\nu}\,.
\]
As was explained in~\cite{Kim:2013zha}, this is the natural off-shell extension of the on-shell ADT current, which leads to the on-shell ADT potential in Einstein gravity.
One of the essential ingredients in the off-shell conservation of this current is the off-shell identity $\nabla_{\mu}(\CG^{\mu\nu}\xi_{\nu})=0$ for a Killing vector $\xi$.  By using the off-shell identity   given in Eq.~(\ref{idaaa}), one can see that   for a Killing vector $\xi$ there is an analogous identity even in the presence of arbitrary matters in the form of 
\begin{equation} \label{ADTcon1}
\nabla_{\mu}({\bf E}^{\mu\nu}\xi_{\nu})=0\,. 
\end{equation}

Inspired by this observation, we introduce the off-shell ADT current for a Killing vector $\xi$ in the presence of arbitrary matter fields  by
\begin{equation} \label{ADT}
\CJ^{\mu}_{ADT} (\xi, \delta \Psi) = \delta {\bf E}^{\mu\nu}\xi_{\nu} + \frac{1}{2}g^{\alpha\beta}\delta g_{\alpha\beta}\, {\bf E}^{\mu\nu}\xi_{\nu}  + {\bf E}^{\mu\nu}\delta g_{\nu\rho}\, \xi^{\rho} + \frac{1}{2}\xi^{\mu}\CE_{\Psi}\delta \Psi\,,
\end{equation}
or, more compactly, in the form of
\begin{equation} \label{ADTcpt}
\sqrt{-g}\CJ^{\mu}_{ADT} (\xi, \delta \Psi) = \delta\Big(\sqrt{-g}\, {\bf E}^{\mu\nu}\xi_{\nu}\Big) + \frac{1}{2}\sqrt{-g}\,\xi^{\mu}\CE_{\Psi}\delta \Psi\,,
\end{equation}
where $\delta$ denotes the generic variation of $\Psi$ such that $\delta \xi^{\mu}=0$.  %
We would like to emphasize that the above construction of the off-shell ADT current, $\CJ_{ADT}$, depends only on the Euler-Lagrange expressions of metric and matter fields. 
  Using this result,  it is straightforward to show the identical conservation of the above off-shell ADT current for a Killing vector $\xi$ in the presence of matter fields.  
The identical conservation of the off-shell ADT current even in the presence of matter fields allows us to introduce the off-shell ADT potential $Q^{\mu\nu}$ as 
\begin{equation} \label{}
\CJ^{\mu}_{ADT} =  \nabla_{\nu}Q^{\mu\nu}_{ADT}\,.
\end{equation}
%


\subsection{Comparison with the covariant phase space approach}

We would like to connect the off-shell ADT current in the presence of matter fields to the symplectic current in the covariant phase space approach~\cite{Lee:1990nz,Wald:1999wa}. To this purpose, it is very useful to introduce the off-shell Noether current. For the simplicity of the presentation, let us focus on the action without gravitational Chern-Simons terms in the following. This means that we are considering the case of $\Sigma^{\mu}=0$ in Eq.~(\ref{Diffeo}).  Recall that the Lagrangian transforms under the diffeomorphism as 
\begin{equation} \label{DiffeoL}  
\delta_{\zeta}  (\sqrt{-g}\CL)= \sqrt{-g}\,\CE_{\Psi}\Lie_{\zeta} \Psi + \p_{\mu}\Theta^{\mu}(\Lie_{\zeta} \Psi)\,.
\end{equation}
Then, one can deduce that the off-shell Noether current in the presence of matter fields can be introduced as
\begin{equation} \label{offNoe}
J^{\mu}(\zeta) = 2\sqrt{-g} {\bf E}^{\mu\nu}\zeta_{\nu} +  \zeta^{\mu}\sqrt{-g}\,\CL - \Theta^{\mu}(\Lie_{\zeta}g, \Lie_{\zeta}\psi)\,.
\end{equation}
%
In order to check that  the identical conservation of this current\footnote{For another direction to use off-shell currents, see \cite{Padmanabhan:2013lpa}.},  
we equate two forms of the diffeomorphism variation  given in Eq.~(\ref{Diffeo}) and in Eq.~(\ref{DiffeoL}) and use the  off-shell identity given in Eq.~(\ref{idaaa}).
%
From these one can  confirm that
$\partial_\mu J^\mu=0$, identically.
Note that the above off-shell Noether current reduces to the on-shell  one by using the EOM of metric and matter fields, $ {\bf E}^{\mu\nu}=0$.  
%
%
The conservation of the off-shell Noether current $J^{\mu}$ allows us to introduce the off-shell Noether potential $K^{\mu\nu}$ as 
\begin{equation} \label{}
J^{\mu} \equiv \p_{\nu}K^{\mu\nu}\,.
\end{equation}

Now, let us recall that symplectic current in the covariant phase space formalism is introduced as~\cite{Lee:1990nz}
\begin{equation} \label{}
\omega^{\mu}(\delta_1 \Psi \,,\, \delta_2\Psi) \equiv \delta_1\Theta^{\mu}(\delta_2 \Psi )  - \delta_2\Theta^{\mu}(\delta_1 \Psi )  \,.
\end{equation}
By using the generic variation of the Lagrangian, the Lie derivative of the surface term 
\[
\Lie_{\zeta}\Theta^\mu(\delta \Psi) = \zeta^{\nu} \partial_{\nu} \Theta^{\mu} - \Theta^{\nu} \partial_{\nu} \zeta^{\mu} + \Theta^{\mu}\partial_{\nu} \zeta^{\nu}\,,
\]
and the invariance property of the diffeomorphism parameter  under a generic variation, $\delta \zeta^{\mu}=0$,
we have 
\begin{equation} \label{Sangb}
\zeta^{\mu}\sqrt{-g}\,\CE_{\Psi}\delta \Psi  = \delta\Big(\zeta^{\mu}\sqrt{-g}\CL\Big) - \p_{\nu}\Big(2\zeta^{[\mu}\Theta^{\nu]}(\delta \Psi)\Big) - \Lie_{\zeta}\Theta^{\mu}(\delta \Psi)\,.
\end{equation}
By varying Eq.~(\ref{offNoe}) and using Eq.s (\ref{ADTcpt}) and (\ref{Sangb}), we obtain one of our essential results:
\begin{equation} \label{Currelation}
2\sqrt{-g}\CJ^{\mu}_{ADT}(\zeta, \delta \Psi) = \p_{\nu}\Big(\delta K^{\mu\nu}(\zeta) - 2\zeta^{[\mu}\Theta^{\nu]}(\delta \Psi) \Big) -\omega^{\mu}(\Lie_{\zeta} \Psi\,,\, \delta \Psi)\,. 
\end{equation}
We would like to emphasize that this relation holds for any background field  configuration and any generic variation, since the conservation of the off-shell ADT current does not require the matter EOM neither  the metric EOM.
For a Killing vector $\xi$, the symplectic current vanishes because $\Lie_{\xi}\Psi=0$. As a result, one can see that  off-shell ADT potential for a Killing vector $\xi$ is identical with the potential $W^{\mu\nu}$ in the  covariant phase space approach~\cite{Wald:1993nt, Iyer:1994ys} as 
\begin{equation} \label{Relation}   
2\sqrt{-g}\, Q^{\mu\nu}_{ADT} (\xi, \delta \Psi) = \delta K^{\mu\nu}(\xi) - 2 \xi^{[\mu} \Theta^{\nu]}(\delta \Psi) \equiv W^{\mu\nu}(\xi, \delta \Psi)\,.
\end{equation}
This proves the complete equivalence between the quasi-local formalism and the covariant phase space approach even in the presence of generic matter fields. 

 In order to obtain finite conserved charges of black holes from  the above ADT potential, we integrate the infinitesimal form of  the potential with respect to  parameters ${\cal Q}_{s}$'s in the  black hole solution, as was adopted in Ref.~\cite{Wald:1999wa,Barnich:2001jy,Barnich:2007bf,Barnich:2003xg,Compere:2007az,Barnich:2004uw}.
Finally, by assuming that the integral is path independent, the finite conserved charge for a Killing vector can be introduced as
\begin{eqnarray}   \label{Finite}
Q(\xi)&\equiv& \frac{1}{8\pi G}  \int ds \int d^{D-2}x_{\mu\nu}\sqrt{-g}\,Q^{\mu\nu}_{ADT} \nn \\
&=&  \frac{1}{16\pi G}\int d^{D-2}x_{\mu\nu} \bigg(\Delta K^{\mu\nu} (\xi)- 2\xi^{[\mu} \int ds~  \Theta^{\nu]}(g\,;\, \CQ_s)\bigg)\,, 
\end{eqnarray}
where $\Delta K^{\mu\nu}$ denotes the finite difference defined by  $\Delta K^{\mu\nu} \equiv K^{\mu\nu}_{{\cal Q}} - K^{\mu\nu}_{{\cal Q}=0}$ and $d^{D-2}x_{\mu\nu}$ denotes the area element of codimension-two subspace.  This final expression of quasi-local conserved charges is completely identical with the one   in the covariant phase space~\cite{Wald:1993nt,Iyer:1994ys} and in the BBC formalism~\cite{Barnich:2001jy,Barnich:2007bf,Compere:2007az}.   This formula can be applied to the computation of the black hole entropy as well as  the mass and angular momentum of black holes. From the properties of the Killing vector on a Killing horizon and the rotational Killing vector,  one can see that the entropy and the angular momentum of black holes can be computed just by the first term in the above formula.

\subsection{Some models}

As an application of our formulation, let us consider the  general two derivative Lagrangian of the form
\begin{eqnarray}   \label{model}
 I  &=& \frac{1}{16\pi G}\int d^Dx~\sqrt{-g} (\CL_g + \CL_{\phi}+\CL_{A} ) 
\end{eqnarray}
where
\begin{align}   \label{model2}
 \CL_{g} =& R - 2\L\,,\qquad
 \CL_{\phi} = - \frac{1}{2}G_{IJ}(\phi)\p_{\mu}\phi^{I} \p^{\mu}\phi^{J} - V(\phi) \,, \qquad
 \CL_{A} = - \frac{1}{4}\CN(\phi)F^{\mu\nu}F_{\mu\nu} \,. 
\end{align}
Explicitly, the variation of the Lagrangian is given by 
\begin{equation} \label{Example}
\delta (\sqrt{-g} \CL) = \sqrt{-g} \Big(\CE_{\mu\nu}\delta g^{\mu\nu} + \CE^{\phi}_{I}\delta \phi^{I} + \CE^{\mu}_{A}\delta A_{\mu} \Big)+\p_{\mu}\Theta^{\mu}\,,
\end{equation}
where the Euler-Lagrange expressions for each field are 
\begin{eqnarray}  
\CE_{\mu\nu} &\equiv & G_{\mu\nu} ^{\L}- T_{\mu\nu}\,,   \qquad \CE^{\nu}_{A} \equiv \nabla_{\mu}(\CN F^{\mu\nu})\,, \label{ExEOM} \\
  \CE^{\phi}_{I}  &\equiv&  G_{IJ}(\phi) (\nabla^2\phi^{J} + \Gamma^{J}_{KL}\p_{\mu}\phi^K\p^{\mu}\phi^L) - \p_{I}V(\phi) - \frac{1}{4}\p_{I}\CN\, F_{\mu\nu}F^{\mu\nu}\,, \nn 
\end{eqnarray}
and the surface terms are given by
\begin{eqnarray}  \label{surfaceS}
  \Theta^{\mu}(\delta g, \delta \phi,\delta A) &=&  \Theta^{\mu}_{g} (\delta g)+   \Theta^{\mu}_{\phi} (\delta \phi) +   \Theta^{\mu}_{A} (\delta A) \\
  &=&   \sqrt{-g} \left[ \,2 g^{\alpha[\mu}\nabla^{\beta]}\delta g_{\alpha\beta}     - G_{IJ}(\phi)\delta\phi^{I} \p^{\mu}\phi^{J} - \CN F^{\mu\nu}\delta A_{\nu}\right]\,. \nn
\end{eqnarray}
Here, Einstein and bulk stress tensors become
\begin{eqnarray}
G_{\mu\nu}^{\L} &=& R_{\mu\nu} - \frac{1}{2}g_{\mu\nu}R + \L \,g_{\m\n} \,, \nn \\
T^{\phi}_{\mu\nu} &=& \frac{1}{2}G_{IJ}(\phi) \p_{\mu}\phi^{I} \p_{\nu}\phi^{J} + \frac{1}{2}g_{\mu\nu} \CL_{\phi}
\,, \nn \\
T^{A}_{\mu\nu} &=& \frac{1}{2}\CN  F_{\mu\alpha}F_{\nu}^{~\alpha}+ \frac{1}{2}g_{\mu\nu}\CL_{A}\,. \nn
\end{eqnarray}
The metric, scalar and gauge field EOM are given by $\CE^{\mu\nu}=0$,   $\CE^{\phi}_{I}=0$ and  $\CE^{\mu}_{A}=0\,. $

As mentioned earlier, several interesting features appear in  the model with  vector fields.
One may modify the Lie derivative of gauge fields since those fields may be accompanied by a certain gauge transformation. In order to use the off-shell identity for the gauge field, it is more useful to introduce a modified Lie derivative which is augmented by a certain gauge transformation such that 
\[   
\Lie'_{\zeta}A_{\mu} \equiv -F_{\mu\nu}\xi^{\nu} = \Lie_{\zeta}A_{\mu} + \p_{\mu}\Lambda\,, \qquad \Lambda \equiv -\zeta^{\alpha}A_{\alpha}\,.
\]
By recalling that gauge fields satisfy a Bianchi identity in the form of 
\[   
\nabla_{[\rho}F_{\mu\nu]}=0\,
\]
and using this modified Lie derivative, one can show that the $\CZ^{\mu\nu}\zeta_{\mu}$  term is absent in Eq.~(\ref{ida}). Surely, this modification is not essential and the unmodified form can also be used without affecting the final result of conserved charges. 
%
%
%
%
For massive gauge fields, one cannot use the modified Lie derivative since there is no gauge invariance. Rather we should keep the original Lie derivatives
\begin{equation} \label{}
\Lie_{\zeta} A_{\mu} = - F_{\mu\nu}\zeta^{\nu} + \p_{\mu}(\zeta^{\nu}A_{\nu})\,.\nn
\end{equation}
In this case,  it turns out that the tensor $\CZ^{\mu\nu}$ is given in terms of the Euler-Lagrange expression $\CE^{\mu}_{A}$ of a gauge field $A_{\mu}$ as 
\begin{equation} \label{}
\CZ^{\mu\nu} = \CE^{\mu}_{A}A^{\nu}\,.
\end{equation}
%
%
%
%
%
%
Just as in the massless case, one can see that the final results on the relation between the off-shell ADT potential and the covariant phase space potential should remain the same as Eq.~(\ref{Relation}).

Now we obtain the ADT potential in this model by using Eq. (\ref{Relation}). Since surface terms are given in Eq. (\ref{surfaceS}), it is sufficient to derive the expression of the Noether potential. By using the following off-shell identities in the Noether current and potential,
\begin{eqnarray}  
 -2\sqrt{-g}\, T^{\mu\nu}_{\phi}\zeta_{\nu} + \zeta^{\mu}\sqrt{-g}\CL_{\phi} - \Theta^{\mu}_{\phi}(\Lie_{\zeta} \phi) &=&0\,,   \\
  -2\sqrt{-g}\, T^{\mu\nu}_{A}\zeta_{\nu} + \zeta^{\mu}\sqrt{-g}\CL_{A} - \Theta^{\mu}_{A}(\Lie'_{\zeta}A) &=& 0\,,
\end{eqnarray}
one can see that the Noether potential  is given by
\begin{equation} \label{NpotEin}
K^{\mu\nu}(\xi) = 2\sqrt{-g}\nabla^{[\mu}\xi^{\nu]}\,. 
\end{equation}
This Noether potential as well as 
the corresponding off-shell Noether current $J^{\mu}$, even  in the presence of matter fields,  take the identical forms with those without matter fields. 
The form of the Noether potential in Eq.~(\ref{NpotEin}) explains why there is no apparent contribution  of matter fields on the entropy of  charged black holes in Einstein gravity, and thus it is simply determined  by the area law. As is well-known, the Wald's  entropy of black holes is captured by the Noether potential only since the contribution of the surface term in $W^{\mu\nu}$, in Eq.~(\ref{Relation}), vanishes on a Killing horizon.  In other words,  any contribution of matter fields to the black hole entropy should be indirectly incorporated through the back reaction of the metric due to matter fields.

In this model, the total off-shell ADT potential is given by the sum of the metric, scalar and gauge field contributions as
\begin{equation} \label{MatterCon}
Q^{\mu\nu}_{ADT} (\xi\,;\, \delta \Psi) = Q^{\mu\nu}_{ADT} (\xi\,;\, \delta g)
 + Q^{\mu\nu}_{ADT} (\xi\,;\, \delta \phi^{I} )  +  Q^{\mu\nu}_{ADT} (\xi\,;\, \delta A )~. 
\end{equation}
By using our relation~(\ref{Relation}), one can easily show that,
for a Killing vector $\xi$, the metric contribution to the off-shell ADT potential  is given by
\begin{equation} \label{MetricCon}
 Q^{\mu\nu}_{ADT} (\xi\,;\,\delta g) =  -\frac{1}{2}g_{\a\b}\d g^{\a\b} \nabla^{[\mu}\xi^{\nu]}    + \xi^{[\mu} \nabla_{\alpha} \d g^{\nu]\alpha} - \xi_{\alpha}\nabla^{[\mu}\d g^{\nu]\alpha}  - g_{\a\b}\xi^{[\mu}\nabla^{\nu]} \d g^{\a\b}  + \d g^{\alpha [\mu} \nabla_{\alpha} \xi^{\nu]}\,,
\end{equation}
and the contributions from the scalar and gauge fields are  given by
\begin{equation} \label{MatterADT}   
Q^{\mu\nu}_{ADT}(\xi\,;\, \delta \phi ) =  G_{IJ}(\phi)\delta\phi^{I} \xi^{[\mu} \p^{\nu]}\phi^{J}\,, \qquad Q^{\mu\nu}_{ADT} (\xi\,;\, \delta A ) =  \CN\, \xi^{[\mu} F^{\nu]\alpha}\delta A_{\alpha}\,.
\end{equation}

Traditionally, matter contributions through $Q^{\mu\nu}_{ADT}$ to total conserved charges have been ignored by supposing that  matter fields  fall off fast when they approach to the asymptotic infinity. However, one need to incorporate those with the slow falloff boundary condition, especially  in the context of the AdS/CFT correspondence since matter contributions have some dual interpretation.

%
%

\section{ Quasi-local formalism and boundary stress tensor method}

In this section we introduce the boundary off-shell current  according to the spirit of our bulk construction and compare conserved charges by this current with those from the bulk off-shell ADT potential.   
In the context of the AdS/CFT correspondence there is another way to obtain conserved charges from the renormalized boundary stress tensor. 
We show that the construction of our boundary current is a kind of the reformulation of  the conventional boundary stress tensor method along our bulk construction. Furthermore,  we show that  conserved charges by our boundary current or from boundary stress tensor method match completely with those from the bulk ADT formalism. 

\subsection{The boundary off-shell current} 

For the construction of the boundary current in the asymptotic AdS space, let us recall that ADM decomposition along the radial direction can be taken as
\begin{equation} \label{}
ds^2 =g_{\mu\nu}dx^{\mu}dx^{\nu} =  N^2dr^2 + \gamma_{ij}(r,x)(dx^{i}+N^idr)(dx^{j}+N^jdr)\,, 
\end{equation}
where $i,j=0,1,\cdots, D-2$. In the following, we denote the space-time dimension of the dual field theory as $d\equiv D-1$. 
To obtain conserved charges from the holographic renormalization perspective~\cite{Henningson:1998gx,deBoer:1999xf,deHaro:2000xn,Bianchi:2001kw,deBoer:2000cz,Skenderis:2002wp}, one may consider the renormalized   action which includes the GH boundary term  $I_{GB}$ and the counter term $I_ {ct}$ as
\[   
I_{r}[g, \psi] = I [g, \psi]+ I_{GH}[\gamma] + I_{ct}[\gamma,\psi]\,,
\]
where the GH boundary and counter terms are defined on a hypersurface  and depends on the boundary values of $\gamma$ and $\psi$ there. The on-shell valued renormalized action $I_{r}^{on}$ would be the functional of the boundary value $(\gamma, \psi)$ at the boundary $\CB$. The generic variation of the on-shell renormalized action is given by\footnote{Our  convention for the boundary stress tensor $T^{ij}_{B}$ is such that it denotes only the finite part after  holographic renormalization and thus corresponds  to $\pi^{ij}_{\textrm{\tiny (d)}}$ in Ref.~\cite{Papadimitriou:2005ii}. And so is the matter part $\Pi_{\psi}$ .} 
\begin{equation} \label{ReAction}
\delta I^{on}_{r}[\gamma, \psi] = \frac{1}{16\pi G}\int_{\CB} d^{d} x~ \sqrt{-\gamma}\Big[ T^{ij}_{B}\delta \gamma_{\,ij} + \Pi_{\psi}\delta \psi\Big]\,,
\end{equation}
where the boundary stress tensor, up to the radial rescaling, $T^{ij}_{B}$ is 
%
%
identified with the stress tensor of dual CFT according to the AdS/CFT correspondence and the renormalized momentum $\Pi_{\psi}$ of the matter field $\psi$ corresponds to  the vev of the  operator dual to the matter field.

One can construct the identically conserved boundary current $\CJ^{i}_{B}$ from the on-shell renormalized action. We begin with  the  identity,  analogous to the bulk one given in Eq.~(\ref{Bulkid}),
\begin{equation} \label{Biden}
-2\zeta_{j}\nabla_{i}T^{ij}_{B} + \Pi_{\psi}\Lie_{\zeta} \psi= \nabla_{i}(\CZ^{ij}_{B}\zeta_{j})\,,
\end{equation}
where  $\zeta$ denotes an arbitrary boundary diffeomorphism paramter and $\CZ^{ij}_{B}$ tensor is a certain combination of $\Pi_{\psi}$. 
This identity follows from  the boundary  diffeomorphism invariance. Just as in the bulk case, the scalar field contribution to $\CZ^{ij}_{B}$ tensor vanishes generically and the vector field contribution to $\CZ^{ij}_{B}$ tensor turns out to be  given by $\Pi^i_{A}A^{j}$. 
Then one can introduce the boundary ADT-like current for a boundary Killing vector $\xi_B$ as
\begin{equation} \label{}
 \CJ^{i}_{B} (\xi_B) \equiv \, - \,\delta{\bf T}^{ij}_{B}  \xi^{B}_{j} - \frac{1}{2}\gamma^{kl}\delta\gamma_{kl} {\bf T}^{ij}_{B}  \xi^{B}_{j} - {\bf T}^{ij}_{B}  \delta\gamma_{jk}\xi_{B}^{k} + \frac{1}{2}\, \xi^{i}_{B}\Big(T^{kl}_{B}\delta \gamma_{kl} + \Pi_{\psi}\delta \psi\Big)
\end{equation}
where 
\begin{equation} \label{}
{\bf T}^{ij}_{B} \equiv T^{ij}_{B} + \frac{1}{2}\CZ^{ij}_{B}\,.
\end{equation}
By using $\delta\xi_{B}^{i}=0$, this boundary current can be written more compactly as
\begin{equation} \label{BdADTcur}
\sqrt{-\gamma} \CJ^{i}_{B} (\xi_B) = -\,  \delta \Big(\sqrt{-\gamma}\, {\bf T}^{ij}_{B}  \xi^{B}_{j}\Big) + \frac{1}{2}\sqrt{-\gamma}\, \xi^{i}_{B}\Big(T^{kl}_{B}\delta \gamma_{kl} + \Pi_{\psi}\delta \psi\Big)\,.
\end{equation}
The above boundary current  takes the analogous form of the bulk off-shell ADT current given in Eq.~(\ref{ADTcpt}) except for the absence of a generalized Einstein tensor. This is natural since the boundary metric field is non-dynamical.  By using   the fact that  %
\beq
\nabla_{i}({\bf T}^{ij}_{B}\xi^{B}_{j})=0\,, 
\eeq
for the boundary Killing vector $\xi_B$, 
one can show that the corresponding current, $\CJ_B^i$, is also conserved identically for a generic variation such that $\delta \xi^i_{B}=0$.  Note that one may regard $\CJ_{B}^{i}$ as a 1-form on the solution parameter space. In order to introduce boundary conserved charges,  we integrate  the 1-form  boundary current in the same manner as in the bulk case. Therefore the boundary conserved charges are given by 
\begin{equation} \label{}
Q_{B}(\xi_B) = \frac{1}{8\pi G}\int_{\partial\CB} d^{d-1}x_i\, \int ds~ \sqrt{-\gamma}\,  \CJ^{i}_{B}(\xi_{B})\,,
\end{equation}
where we integrate over the path parametrized by $s$ in the  parameter space in the given solution\footnote{As a working hypothesis,  we assume that the 1-form boundary current is  independent of path. This assumption holds in all the examples given in the following sections.}.

\subsection{The equivalence with the boundary stress tensor method}

We would like to uncover the relation between  (linearized) conserved charges obtained from the boundary current introduced in the previous section and those from  the conventional boundary stress tensor method in
~\cite{Balasubramanian:1999re, Papadimitriou:2005ii,Hollands:2005wt}.  
As alluded earlier,  we perform the variation $\delta$ along the one-parameter path in the solution space. 
As will be explained through examples,  the one-parameter path in the solution space corresponds to the choice of   a representative in the conformal class at the boundary with a restricted diffeomorphism preserving the gauge choice. 

When  the contribution from the second term of the boundary current $ \CJ^{i}_{B} $  in Eq.~(\ref{BdADTcur}) is absent,  
the boundary current reduces to 
\begin{equation} \label{BCurr}
\sqrt{-\gamma} \CJ^{i}_{B} = - \delta \Big(\sqrt{-\gamma}\, {\bf T}^{ij}_{B}\xi^{B}_{j}\Big) \,.
\end{equation}
By using the conventional expression of holographic charges in the form of
\begin{equation} \label{qqqqq}
\hat{Q}_{B}(\xi_B) = -\frac{1}{8\pi G}\int_{\partial\CB} d^{d-1}x_i~ \sqrt{-\gamma}\, {\bf T}^{ij}_{B}\, \xi^{B}_{j}  \,,
\end{equation}
the expression of finite conserved charges for the Killing vector $\xi_{B}$ from the boundary ADT formalism can be obtained as 
\begin{equation} \label{qqq}
Q_{B}(\xi_B) = \hat{Q}_{B}(\xi_{B}) - \hat{Q}^{AdS}_{B}(\xi_{B})\,.
\end{equation}
This verifies the equivalence,  up to the AdS vacuum value, between  the boundary quasi-local ADT formalism and  the conventional boundary stress tensor method. %

In order to see the meaning of the second term in Eq. (\ref{BdADTcur}), let us focus on the specific model introduced in (\ref{model}). 
 In this model, we would like to consider the relation between the allowed boundary condition on the asymptotic AdS space and   the absence of the contribution from the second term of the boundary current $ \CJ^{i}_{B} $.   
As was discussed in Ref.~\cite{Papadimitriou:2005ii} in the context of the well-posedness of the  variational problem, the boundary condition allowed in the asymptotic AdS space  
needs to be relaxed as
\begin{equation} \label{}
\delta \g_{ij} = 2\g_{ij}\delta \sigma\,, \qquad \delta A_{i} =0\,, \qquad \delta \phi^I = (\Delta_I - d)\phi^I\delta \sigma\,,
\end{equation}
where $\Delta_{I}$ is the conformal dimension of dual operator to a scalar field $\ph^{I}$. This boundary condition shows us that  the second term in Eq.~(\ref{BdADTcur}) is nothing but the conformal anomaly $\CA$ in the boundary field theory.  Explicitly, the second term becomes
\begin{equation} \label{}
T^{kl}_{B}\, \delta \gamma_{kl} + \Pi_{\psi}\, \delta \psi = \Big[2\,T^{i}_{B~ i} + \sum_{I}(\Delta_I - d)\Pi_{\phi^I} \phi^I\Big]\delta \sigma \equiv \CA\delta \sigma\,.
\end{equation}

There is no conformal anomaly  in the dual field theory of the even dimensional AdS geometry. On the other hand, in odd dimensional AdS geometry the dual CFT has conformal anomaly.   
We consider the boundary conditions of metric and matter fields satisfying  $\int \delta \sigma\, \CA =0$, which holds in all our examples. This leads to  the absence of the contribution from the second term in the boundary current in Eq. (\ref{BdADTcur}).

Due to the the absence of the scalar field contribution to $\CZ^{ij}_{B}$, we have 
\begin{equation} \label{}
{\bf T}^{ij}_{B} = T^{ij}_{B} + \frac{1}{2} \Pi^{i}_{A}A^{j}\,,
\end{equation}
and we can see that  Eq. (\ref{qqq}), up to AdS vacuum value, gives us the identical expression of  conserved holographic charges with the one in the conventional boundary stress tensor method (See the Eq.~(4.28) in Ref.~\cite{Papadimitriou:2005ii}).

\subsection{The equivalence with the bulk ADT potential}

In this section we would like to  show that the boundary current and the bulk potential lead to the same conserved charges. One may recall that the holographic renormalization process introduces new boundary terms in the given Lagrangian with the on-shell condition. These new boundary terms do not affect the bulk EOM, and thus the construction of the bulk current given in Eq.~(\ref{ADT}), which depends only on the bulk Euler-Lagrange expressions,  is valid and so can be used  without any modification. The effect of the new boundary terms comes in through the modifications of the Noether potential $K^{\mu\nu}$ and the surface term $\Theta^{\mu}$ given in Eq.~(\ref{Relation}).

For definiteness,  it is convenient to use the, so-called, Fefferman-Graham(FG) coordinates for an asymptotically AdS space~\cite{Fefferman} which is given in the form of 
\begin{align}   \label{FGexp}
ds^2 = d\eta^2 + \gamma_{ij}dx^idx^j\,. 
\end{align}
In the following, we take  the radius of asymptotic AdS space unity and the cosmological constant $\L =  -\frac{{d(d-1)} }{2}$. 
In these coordinates the boundary is located at $\eta_0$, which will be sent to be infinity in the end. The radial expansion of the metric and  the matter fields are generically taken as
\begin{align}   \label{}
\gamma_{ij} =  e^{2\eta}\Big[\g^{(0)}_{ij} + \CO(e^{-\eta})\Big]\,, \qquad \qquad
\psi = e^{-(d_\psi -\Delta_{\psi})\eta}\Big[\psi_{(0)} + \CO(e^{-\eta})\Big]\,, 
\end{align}
where $\Delta_{\psi}$  is the conformal dimension of the operator dual to $\psi$ and $d_\psi$ is given by $d_{\psi}=d-p$ for the rank $p$ tensor field $\psi$. The boundary metric $\g^{(0)}_{ij} $ represents the background geometry of the dual CFT according to the AdS/CFT dictionary. Formally 
the GH boundary term and counter term are taken by
\begin{equation} \label{}
I_{GH}[\gamma] = \frac{1}{8\pi G}\int d^{d}x\sqrt{-\gamma}L_{GH}(\gamma)\,, \qquad  I_{ct}[\gamma,\psi] = \frac{1}{16\pi G}\int d^{d}x\sqrt{-\gamma}L_{ct}(\gamma, \psi)\,, 
\end{equation}
which  make the renormalized action finite in the limit $\eta_{0}\rightarrow\infty$.

The modification in boundary terms can be succinctly captured  by the introduction of a modified surface term $\tilde{\Theta}^{\eta}$  as 
\begin{eqnarray}   \label{}
\tilde{\Theta}^{\eta}(\delta \Psi) &=& \Theta^{\eta}(\delta \Psi) + \delta(2\sqrt{-\gamma}L_{GH}) + \delta(\sqrt{-\gamma}L_{ct})  \\
&=& \sqrt{-\gamma}\Big(T^{ij}_{B}\delta \gamma_{ij} + \Pi_{\psi}\delta \psi\Big)\,,  \nn 
\end{eqnarray}
where the second line equality  comes from Eq.~(\ref{ReAction}). 
This expression tells us that $\tilde{\Theta}^{\eta} \sim \CO(1)$ in the radial expansion.
Correspondingly, the modified Noether current $\tilde{J}^{\eta}$ for a diffeomorphism parameter $\zeta$ becomes
\begin{equation} \label{ModRel}
\tilde{J}^{\eta} = \p_{i}\tilde{K}^{\eta i}(\zeta) = \zeta^{\eta}\sqrt{-\gamma}\CL^{on}_{r} - \tilde{\Theta}^{\eta}(\Lie_{\zeta} \Psi)\,,
\end{equation}
where we have used the on-shell condition on the background fields in Eq. (\ref{offNoe}). Here, one may also note that the on-shell renormalized Lagrangian $\sqrt{-\gamma}\CL^{on}_{r}$ is related to the so-called $A$-type trace anomaly~\cite{Deser:1993yx, Henningson:1998gx}. 

Just as in Einstein gravity~\cite{Papadimitriou:2005ii},  the asymptotic behavior of general diffeomorphism parameter $\zeta$  is given by
\begin{equation} \label{}
\zeta^{\eta} \sim \CO(e^{-d\eta})\,, \qquad \zeta^{i} \sim \CO(1)\,,
\end{equation}
in order to preserve the asymptotic gauge choice and the renormalized action.
This asymptotic behavior in the diffeomorphism parameter $\zeta$ allows us to discard the first term in the right hand side of Eq.~(\ref{ModRel}) when we approach the boundary. In the following we keep only the relevant boundary values of parameters such that  a bulk Killing vector $\xi^{i}$ is replaced by its boundary value $\xi^{i}_{B}$.
For the diffeomorphism variation $\Lie_{\zeta}\Psi$, the modified surface term $\tilde{\Theta}^{\eta}$ is given by
\begin{equation} \label{}
\tilde{\Theta}^{\eta}(\Lie_{\zeta}\Psi) = \sqrt{-\gamma}\Big(2T^{ij}_{B}\nabla_{i}\zeta_{j} + \Pi_{\psi}\Lie_{\zeta}\psi\Big) = \p_{i}\Big(2\sqrt{-\gamma}\,{\bf T}^{ij}_{B} \,\zeta_j\Big)\,,
\end{equation}
where we have used the identity given in Eq.~(\ref{Biden}).  By using this result, one can see  that the Noether potential $\tilde{K}^{\eta i}$ becomes 
\begin{equation} \label{}
\tilde{K}^{\eta i} = -2\sqrt{-\gamma}\,{\bf T}^{ij}_{B} \,\zeta_j + \p_{j}(\sqrt{-\gamma}\,\CU^{ij}_{B})\,,
\end{equation}
where $\CU_{B}^{ij}$ is an arbitrary anti-symmetric second rank tensor.
Since we are interested in conserved charges, the total derivative term $\p_{j}(\sqrt{-\gamma}\,\CU^{ij}_{B})$ is irrelevant and can be discarded for simplicity. 
As a result, the relation between the ADT and Noether potentials in Eq.~(\ref{Relation}) for a Killing vector $\xi$ becomes 
\begin{equation} \label{equivalence}
2\sqrt{-g}Q^{\eta i}_{ADT}|_{\eta\rightarrow \infty} = - \delta \Big(2\sqrt{-\gamma}\,{\bf T}^{ij}_{B} \,\xi^{B}_j \Big) + \sqrt{-\gamma} \,\xi^{i}_{B}\Big(T^{kl}_{B}\delta \gamma_{kl} + \Pi_{\psi}\delta \psi\Big) \equiv 2\sqrt{-\gamma}\CJ^{i}_{B}\,.
\end{equation}
That is to say the leading parts of the bulk ADT potential and the boundary current are identical when we go to the asymptotic infinity.\footnote{The holographic charges from  boundary stress tensor method are defined by  the first term only.  In Einstein gravity it was shown in \cite{Papadimitriou:2005ii} that the holographic charges are identical with those from the covariant phase space formalism when conformal anomaly is absent. Our modification of the holographic charges, in which the second term is naturally incorporated,  maintain the equivalence between the holographic and bulk charges. }
This proves the equivalence of conserved charges by the bulk potential, $Q$ and those by the boundary current, $Q_{B}$:
\begin{eqnarray}   \label{}
Q(\xi) = \frac{1}{8\pi G}\int_{\CB} d^{D-2}x_{\eta i} \int ds \sqrt{-g}Q^{\eta i}_{ADT} = \frac{1}{8\pi G}\int_{\partial\CB} d^{d-1}x_i\, \int ds~ \sqrt{-\gamma}\,  \CJ^{i}_{B} = Q_{B}(\xi_{B})\,.
\end{eqnarray}

Our results extend, to a general theory of gravity, the equivalence statement given for a specific model in Ref.~\cite{Papadimitriou:2005ii} and are completely consistent with the rather formal argument on such equivalence given in Ref.~\cite{Hollands:2005ya}. 
 We would like to emphasize that the matching between the ADT potential and the boundary current  is valid only at the boundary,  while the bulk ADT potential in the quasi-local sense  could be applied even to the deep interior region like the black hole horizon.  

\section{Generalities for scalar fields}
In this section we introduce the radial expansion of the metric and matter fields and explain some properties related to the computation of  conserved charges. We also explain how to construct the boundary stress tensor.
For simplicity, we consider only a scalar field in the matter sector with the action  given in Eq.~(\ref{Example}). The boundary metric is taken to be flat  as  $\g^{(0)}_{ij}=\eta_{ij}$. 
In pure Einstein gravity,
the conformal anomaly of the dual field theory is absent as a consequence of the flat boundary metric.  And thus logarithmic terms do not appear in  the metric  and 
 the radial expansion of the on-shell metric, in the  FG coordinates, is generically given by 
 \begin{equation}
\g_{ij} = e^{2\eta} \left(\eta_{ij} +  e^{-d\eta} \g_{ij}^{ (d)} +\cdots \right)\,.
\label{SangD}
\end{equation}
It is well-known that the leading order term,   $e^{-d\eta} \g_{ij}^{ (d)}$,  gives  the well-defined, finite, total conserved charges, like the mass and angular momentum of black holes.

\subsection{The radial expansion}
We assume the scalar field depends only on the radial coordinate $\eta$. In general the leading order in the radial expansion  of  the scalar field is given by $\phi\sim e^{-(d-\Delta_\pm)\eta}\phi_\pm$, where  $\phi_+ $ and  $\phi_-$ correspond to the leading order terms of the non-normalizable and normalizable modes, respectively, and  $\Delta_\pm=\frac{d}{2}\pm \sqrt{ \frac{d^2}{4}+m^2}$. The mass of the scalar field has unitary bound or the Breitenlohner-Freedman(BF)  bound~\cite{Breitenlohner:1982jf}:  $m^2 =m^2_{BF} = -\frac{d^2}{4}$, in which  the exponents degenerate as $ \Delta_+ = \Delta_-=\frac{d}{2}$. In this case the scalar field include the logarithmic mode behaving as $\phi\sim \eta e^{-\frac{d}{2}\eta} \phi_{\rm log}$. 
We consider the BF-saturated case, first.
 
 \vskip0.5cm
 
 {\underline {\it  Class I}} :  $\qquad  m^2 =m^2_{BF} = -\frac{d^2}{4}$
 \vskip0.5cm

We can apply our formalism to the case with the logarithmic mode, which was studied in~\cite{Henneaux:2004zi,Henneaux:2006hk} by using the Hamiltonian formalism. For simplicity, we consider the case in which  the leading order term in the radial expansion starts at the order $e^{-\frac{d}{2}\eta}$ and take the radial expansion as  
\begin{align}   \label{}
\phi &=  e^{-\frac{d}{2}\eta}\Big(\phi_{\textrm{\tiny (0)}} +  \cdots\Big)\,.
\end{align}
The corresponding  radial expansion of the metric solution takes the same form given in Eq.~(\ref{SangD}). 

Now, let us perform a linearized analysis to see the back reaction of the metric to the scalar field. 
By taking into account the leading order behavior of the scalar field, it is sufficient to  take the scalar potential up to quadratic order as
\begin{align}   \label{}
V(\phi)&=  \half m^{2} \phi^{2} + \cdots \,.  
\end{align}
The linearized EOM of our specific model become
\begin{align}
& {h}_{ij}^{\prime\prime}+(d-4) {h}_{ij}^{\prime} + (4-2d) h_{ij} - e^{2\e}\e_{ij} \left( h^{\prime\prime} + d h^{\prime} \right) =0 \,, \\
& (d-1) h' - \frac{d^2}{4}e^{-d \eta} \phi_{\textrm{\tiny (0)}}^2 =0\,, \qquad h \equiv e^{-2\eta} \eta^{ij}h_{ij} \label{EOMaa}\\
& \varphi^{\prime\prime} + d\varphi^{\prime} - m^2 \varphi =0 \,,  
\end{align}
where primes denote derivatives with respect to $\e$ and $ \g_{ij} \equiv e^{2\e}\e_{ij}+h_{ij} ~$  and $\phi \equiv e^{-\frac{d}{2} \eta} \phi_{\textrm{\tiny (0)}}+ \varphi \,.$
 Since the leading order contribution of the scalar field  to the metric starts from the order $e^{-d\eta}$,  the linear analysis is sufficient to compute  conserved charges.  From Eq.~(\ref{EOMaa}), the leading order coefficient $\g_{ij}^{ (d)}$ in metric  satisfies the trace relation, 
\begin{equation}
 \qquad \eta^{ij} \g_{ij}^{ (d)} = - \frac{d}{4(d-1)}  \phi_{\textrm{\tiny (0)}}^2 \,.
 \end{equation}
 The form of the coefficients $\gamma_{ij}^{(d)}$ would be further specified by the metric ansatz of the solution. 
As in the case of pure Einstein gravity, these coefficients can be used to determine  the conserved charges.
 \vskip0.5cm
 {\underline {\it  Class II}} :  $\qquad  m^2 >m^2_{BF} = -\frac{d^2}{4}$
 \vskip0.5cm

In this class we consider the case with $\D_{\phi} = \D_{+}$ and then the radial expansion  of  the scalar field solution  is given in the form of 
\begin{align}   \label{MatSolNonSat}
\phi &= e^{-(d-\Delta_{\phi})\eta}\Big(\phi_{\textrm{\tiny (0)}} +e^{-2(d-\Delta_{\phi})\eta} \phi_{\textrm{\tiny (2)}}+e^{-4(d-\Delta_{\phi})\eta} \phi_{\textrm{\tiny (4)}}+  \cdots\Big)\,,
\end{align}
for the even scalar potential whose generic expansion is given by
\begin{align}   \label{potential}
V(\phi)&=  \half m^{2} \phi^{2} +\frac{1}{4} \l \phi^{4}+  \cdots \,.  
\end{align}
If $\Delta_{\phi}\geq d $, the presence of this non-normalizable mode change the asymptotic AdS structure. Henceforth, we restrict ourselves to the case $\Delta_{\phi}<d$, which corresponds to $m^2<0$.  
The corresponding metric solution  has the radial expansion,
\begin{equation}\label{MetSolNonSat}
\g_{ij} = e^{2\eta} \left[\eta_{ij} +  e^{-2(d-\D_{\phi}) \, \eta}\g_{ij}^{(2d-2\D_\phi)}  + \cdots + e^{-d\eta} \g_{ij}^{ (d)} +\cdots \right]\,,
\end{equation}
where the leading order term in the expansion of the metric is given by  
\begin{equation}
 \g_{ij}^{(2d-2\D_\phi)} = - \frac{  \phi_{\textrm{\tiny (0)}}^2 }{4(d-1)} \eta_{ij} \,.
 \end{equation}

The slower falloff terms than $e^{-d\eta} \g_{ij}^{ (d)}$ may give divergent contributions to conserved charges.  However such divergencies should be automatically taken care and finite values emerge since our bulk formalism, by  using one-parameter path in the solution space, gives  identical results with those from the boundary stress tensor formalism.  One may note that   conserved charges are generically determined by $ \g_{ij}^{ (d)}$. Since the contribution of the scalar source to the metric starts, at least, from the $\phi^2$ term, we need to know all the coefficients up to the order $e^{-(2\Delta_{\phi}-d)\eta}$ in the expansion of the full solution of the scalar field.
This will be clearly shown  through the explicit computation of   conserved charges in specific examples in section 5.

\subsection{Counter terms and boundary stress tensor}
In this section we present the generic forms of the GH and counter terms in the model (\ref{model}).  By using these forms, we give  the resultant form of the boundary stress tensor and the renormalized momentum of the scalar field.
 
First of all, the GH term for the Einstein gravity is given by
\begin{equation} \label{}
L_{GH} = K (\g)  \,,
\end{equation}
where $K(\gamma)$ is extrinsic curvature scalar at the boundary.
The counter terms $L_{ct}(\gamma, \phi)$ consist of two parts,
\begin{equation} \label{}
L_{ct} =2 K_{ct} (\g) +  \Phi_{ct}(\phi) \,,
\end{equation}
where the first term is the counter term for the pure gravity and the second one is the one for the scalar field.
The counter terms for the pure gravity part are given by~\cite{Hyun:1998vg,Balasubramanian:1999re,Mann:1999pc,Emparan:1999pm}
\begin{align}
	K_{ct } (\g)= -(d-1)-\frac{1}{2(d-2)} R_{B} - \frac{1}{2(d-4)(d-2)^2} \left( R^{B}_{ij} R_{B}^{ij} -\frac{d}{4(d-1)} R^{2}_{B} \right) + \cdots \,, 
\end{align}
where $R^{B}_{ij} $ and $R_B$ are intrinsic Ricci tensor and scalar at the boundary, respectively.
The counter terms for the scalar field $\phi$ are chosen  as the polynomial of the scalar field   as
\begin{align}	
\label{phict}	\Phi_{ct}(\phi) =& \alpha_1 \, \phi^2 +  \alpha_2 \, \phi^{4} +  \cdots \,,
\end{align}
where $\alpha_{{k}}$ are determined to cancel the divergences in the renormalized action at the boundary.

It follows that the
boundary stress tensor consists of two parts 
\begin{eqnarray} \label{BEnergy}
T^{ij}_{B} &=& T^{ij}_{g} + T^{ij}_{\phi}\,, 
\end{eqnarray}
where $ T^{ij}_{g}$ and $T^{ij}_{\phi}$ come from the metric and scalar fields, respectively.
They are given by
\begin{eqnarray} \label{BEnergy1}
T^{ij}_{g} &=&K \gamma^{ij} -K^{ij} -{(d-1)} \gamma^{ij} + \frac{1}{(d-2)} \left( R_B^{ij}-\half R_B \g^{ij} \right) +\cdots  \,,  \\
T^{ij}_{\phi} &=&   \frac{\gamma^{ij}}{2} \left( \alpha_1 \, \phi^{2} + \alpha_2\, \phi^4+ \cdots \right)\,. 
\end{eqnarray}
One may note that the contribution of the scalar field to the boundary stress tensor comes only from the counter term action and the concrete expression of $T_B$ depends on the form of the counter term action.
One may also note that  in this case $${\bf T}^{ij}_B=T^{ij}_{B}\,, $$ since we are considering a scalar field only.
The renormalized momentum of the scalar field at the boundary is given by
\begin{equation} \label{}
\sqrt{-\gamma}\,\Pi_{\phi} = \sqrt{-\gamma}\Big[-\p_{\eta}\phi + 2\alpha_1\phi + 4\alpha_2 \phi^{3} + \cdots \Big]\,.
\end{equation}
In class I,    it is sufficient to take $\alpha_1=-\frac{d}{4}$, $\alpha_2=\cdots=0$  and then it turns out  that  $\Pi_{\phi}=0$ generically.

\section{Application to various black holes}
In this section we apply our quasi-local formalism to some specific examples. In particular, we compute total conserved charges from both bulk and boundary constructions. We support the general proof of the equivalence on total charges in the bulk and boundary constructions through explicit computations. 
 All the examples we have presented in this section   correspond  to the  specific cases such that the one-parameter path in the solution space is taken as $\delta_s \gamma^{(0)}_{ij}=0$. 
 In our bulk construction, we compute each contribution from the metric and matter sectors to conserved charges, by using Eqs.~(\ref{MetricCon}) and (\ref{MatterADT}). We find each contribution to conserved charges matches with  the corresponding one in our boundary construction. Specifically we reproduce  the  mass and angular momentum of AdS black holes in various dimensions and explain additional salient features in our formalism through explicit examples.

\subsection{Three-dimensional black holes}
In three-dimensional gravity, we have various analytic black hole solutions which allow us to apply our formalism concretely. 
Specifically, we consider the three-dimensional, AdS black hole space with  scalar hair.

 \vskip0.5cm

 {\underline {\it  Class I}} :  $\qquad  m^2 =m^2_{BF} = -1$
 \vskip0.5cm
By solving the linearized EOM we obtain the most general solution of the metric as
\bea \label{LinSol}
\gamma_{ij}^{(2)} =  \left( \ba{cc}
C_1+\frac{1}{4}\phi_{\textrm{\tiny (0)}}^2& - C_2\\
		 - C_2 &  C_1-\frac{1}{4} \phi_{\textrm{\tiny (0)}}^2
\ea  \right)\,,
\eea
where $C_1$ and $C_2$ are arbitrary parameters which turn out to be proportional to the mass and the angular momentum, respectively, of  AdS black holes with scalar hair.
In order to compute the mass and angular momentum of these black holes  in the bulk quasi-local formalism, 
we take the time-like and rotational Killing vectors  as 
$\xi_T = \frac{\p}{\p t}$ and $\xi_R = \frac{\p}{\p \theta}$ and  take the relevant path in  the solution space parametrized by $C_1, C_2$ and $\phi_{\textrm{\tiny (0)}}$. 

The ADT potentials in Eqs.~(\ref{MetricCon}) and  (\ref{MatterADT}) for the time-like Killing vector $\xi_T^{i}=(1,0)$ are computed as 
\begin{align} \label{} 
\sqrt{-g}\, Q^{\eta i}_{ADT} (\xi_T\,;\, \d g  )\big|_{\eta\rightarrow\infty} &= \Big(\, \d C_1 -\frac{1}{2} \phi_{\textrm{\tiny (0)}} \d \phi_{\textrm{\tiny (0)}}\, , \,\, \d C_2 \,\,\,\Big)  \, ,\\
\sqrt{-g}\, Q^{\eta i}_{ADT} (\xi_T \,;\, \d \phi )\big|_{\eta\rightarrow\infty} &= \Big(~~ ~ \frac{1}{2}\phi_{\textrm{\tiny (0)}} \, \d \phi_{\textrm{\tiny (0)}}~~~\,,\,\,~0~\,\, \Big)\,  
\,.
\end{align}
By using  Eq.~(\ref{Finite}) with the convention  $dx_{\eta t} = \frac{1}{2\sqrt{-g}} \epsilon_{\eta t \theta} d\theta = d\theta$, we obtain
\beq
M_{ADT}^{g}  
= \frac{1}{4 G} \Big( C_1  -\frac{1}{4}\phi^2_{\textrm{\tiny (0)}}\Big) \,, \qquad 
M_{ADT}^\phi   
=  \frac{1}{16G} \phi_{\textrm{\tiny (0)}}^2 \,.
\eeq
Therefore, the total mass of these black holes is  given  by
\begin{align}  M_{ADT} \equiv M^{g}_{ADT} + M^{\phi}_{ADT} = \frac{1}{4G} C_1\,.  
\end{align}                       

The ADT potentials for the rotational Killing vector $\xi_R^{i}=(0,1)$ are computed as
\begin{align} \label{}
\sqrt{-g}& Q^{\eta i}_{ADT}  (\xi_R\,;\, \d g  )\big|_{\eta\rightarrow\infty} \, = \big(-\d C_2 \,\,,\,\, -\d C_1 -\half \phi_{\text{\tiny(0)}}\d \phi_{\textrm{\tiny (0)}}\, \big)  \, ,\\
\sqrt{-g}& Q^{\eta i}_{ADT} (\xi_R\,;\, \d g  )\big|_{\eta\rightarrow\infty} \, =  \big(\,\,\quad 0\,\quad,\,\,\quad \,\, 0 \,\,\,\quad\, \big) \,  
\,.
\end{align}
Therefore the scalar contribution to the angular momentum is absent and the total angular momentum of these black holes is given by
\begin{align}
	J_{ADT}  \equiv J^{g}_{ADT} + J^{\phi}_{ADT}= \frac{1}{4G}C_2\,.
\end{align}   

Now we present the boundary stress tensor  explicitly and confirm the equivalence relation (\ref{equivalence}) between the bulk ADT potential  and the boundary current. After a bit of computation, one obtains the boundary stress tensor as
\bea \label{}
({\bf T}_{g})^{i}\,_{j} =  \left( \ba{cc}
-C_1 + \frac{1}{4} \phi_{\textrm{\tiny (0)}}^2& - C_2\\
		 - C_2 &  C_1+ \frac{1}{4}\phi_{\textrm{\tiny (0)}}^2
\ea  \right)\,, \quad
({\bf T}_{\phi})^{i}\,_{j} =  \left( \ba{cc}
-\frac{1}{4} \phi_{\textrm{\tiny (0)}}^2& 0\\
		 0 &  - \frac{1}{4}\phi_{\textrm{\tiny (0)}}^2
\ea  \right)\,.
\eea
It is straightforward to confirm the equivalence relation (\ref{equivalence}) for Killing vectors $\xi_T$ and $\xi_R$.  One may note that the equivalence relation holds for the metric and matter part separately.

Now we present some known black hole solutions which belong to this class.

\begin{itemize}
\item BTZ black hole solutions~\cite{Banados:1992wn,Banados:1992gq}
\begin{align}
 ds^2 =& - \frac{(r^2 -r_-^2 ) (r^2 - r_+^2)}{r^2} dt^2 + \frac{r^2}{(r^2 -r_-^2 ) (r^2 - r_+^2)}dr^2 + r^2 \left(d\th -\frac{r_- r_+}{r^2 } dt \right)^2\,.
\end{align}
These are solutions in pure gravity with a cosmological constant or  solutions without scalar hair, $ \phi_{\textrm{\tiny (0)}} = 0$.
After transforming to FG coordinates, one can read off
\begin{align}
 C_1 = \frac{r_-^2 + r_+^2}{2} \,, \qquad C_2 = r_- r_+ \,, 
\end{align}
which reproduce the well-known expressions of the total mass and angular momentum of BTZ black holes
\begin{align}
M = \frac{r_-^2 +r_+^2 }{8G}\,, \qquad J= \frac{r_- r_+}{4G} \,.
\end{align}

\item The extremal rotating black holes with scalar hair \cite{Hyun:2012bc,Kwon:2012zh,Hotta:2008xt}
\begin{align}
	\label{}
ds^2 = & \, r^2 \left[-1  + \frac{\mu_0}{r^2} + {\cal{O}}\big(\frac{1}{r^3}\big) \right] dt^2 +\frac{1}{r^2}\left[1 +\frac{\mu_0 - \frac{1}{2}\phi^2_{\textrm{\tiny (0)}}}{r^2}  + {\cal{O}} \big( \frac{1}{r^3} \big) \right]dr^2 \\
	                & + r^2\left[d\theta - \Big(\frac{\mu_0}{2 r^2} + {\cal{O}} \big(\frac{1}{r^3}\big) \Big)  dt\right]^2 \,, \nn                                                                                                              \\
	\phi(r) =       & \,  \frac{\phi_{\textrm{\tiny (0)}}}{r} + {\cal{O}}\big(\frac{1}{r^2}\big)\,,
\end{align}
These are solutions corresponding to the case $C_1 = C_2 = \frac{\mu_0}{2}$. The total mass and  angular momentum of these black holes are computed as
\begin{align}
M = J = \frac{\mu_0}{8G} \,,
\end{align}
which satisfy the extremality condition.

\end{itemize}

 \vskip0.5cm
 {\underline {\it  Class II}} :  $\qquad   -1 < m^2 < 0$
 \vskip0.5cm
In this class we apply our formalism to those solutions given in Ref.~\cite{Henneaux:2002wm,Henneaux:2006hk}. The scalar potential with a cosmological constant is taken as
\begin{equation}
V(\phi) - 2  = -2 \, \Big[ \cosh^6 (\frac{\phi}{4}) + \n \sinh^6 (\frac{\phi}{4})  \Big] \,.
\end{equation}
The radial expansion, Eq.(\ref{MatSolNonSat}), of the scalar field  in FG coordinates  becomes
 \begin{equation}
 \phi =e^{-\frac{1}{2} \eta} \Big(\,  \phi_{\textrm{\tiny (0)}} +\frac{1}{48}\phi_{\textrm{\tiny (0)}}^3 e^{- \eta}+\cdots  \Big) \,,
\end{equation}
 while the coefficients in the radial expansion of the metric solution up to the $e^{-2\eta}$ order are  given by
\bea
\gamma^{(1)} _{ij} =   -\frac{1}{4} \phi_{\textrm{\tiny (0)}}^2 \, \eta_{ij} \,, \qquad \gamma^{(2)}_{ij} = \frac{3}{128} \phi^4_{\textrm{\tiny (0)}}\left[\eta_{ij} +\frac{(1+\nu)}{4} \delta_{ij} \right]\,.
\eea 
The ADT potentials  for the time-like Killing vector $\xi_T^{i}$ are computed  as
\begin{align} \label{} 
\sqrt{-g}\, Q^{\eta i}_{ADT} (\xi_T\,;\, \d g  )\big|_{\eta\rightarrow\infty} &=  \Big[  -\frac{1}{4} e^{\eta}  \phi_{\textrm{\tiny (0)}} \d \phi_{\textrm{\tiny (0)}} +        \frac{1}{32}  \phi_{\textrm{\tiny (0)}}^3 \d \phi_{\textrm{\tiny (0)}} + \frac{3(1+\nu)}{128} \phi_{\textrm{\tiny (0)}}^3 \d \phi_{\textrm{\tiny (0)}}  \, \Big] \,\xi^{i}_{T} \,, \\
\sqrt{-g}\, Q^{\eta i}_{ADT} (\xi_T \,;\, \d \phi )\big|_{\eta\rightarrow\infty} &= \Big[ ~\frac{1}{4} e^{\eta}  \phi_{\textrm{\tiny (0)}} \d \phi_{\textrm{\tiny (0)}}  -\frac{1}{32} \phi_{\textrm{\tiny (0)}}^3 \d \phi_{\textrm{\tiny (0)}} ~ \Big]\, \xi_{T}^{i}\,  
\,.
\end{align}
By using the Eq.~(\ref{Finite}), we obtain the total mass of black holes
\begin{align}  M_{ADT} \equiv  M^{g}_{ADT} +  M^{\phi}_{ADT} =  \frac{1}{4G} \frac{3(1+\nu)}{512} \phi_{\textrm{\tiny (0)}}^4   \,.  
\end{align}                       
The ADT potentials for the rotational Killing vector $\xi_R^{i}$ become
\begin{align} \label{} 
\sqrt{-g}\, Q^{\eta i}_{ADT} (\xi_R\,;\, \d g  )\big|_{\eta\rightarrow\infty} &=  \Big[ - \frac{1}{4} e^{\eta} \phi_{\textrm{\tiny (0)}} \d \phi_{\textrm{\tiny (0)}} +\frac{1}{32} \phi_{\textrm{\tiny (0)}}^3 \d \phi_{\textrm{\tiny (0)}}  -\frac{3(1+\n)}{128} \phi_{\textrm{\tiny (0)}}^3 \d \phi_{\textrm{\tiny (0)}}\, \Big]\, \xi^{i}_{R} \,, \\
\sqrt{-g}\, Q^{\eta i}_{ADT} (\xi_R \,;\, \d \phi )\big|_{\eta\rightarrow\infty} &= \Big[ ~\frac{1}{4} e^{\eta}  \phi_{\textrm{\tiny (0)}} \d \phi_{\textrm{\tiny (0)}}  -\frac{1}{32} \phi_{\textrm{\tiny (0)}}^3 \d \phi_{\textrm{\tiny (0)}} ~ \Big] \, \xi_{R}^{i}\,  
\,.
\end{align}
Therefore it turns out  that the total angular momentum vanishes 
\begin{align} 
                       J_{ADT}  \equiv J^{g}_{ADT} + J^{\phi}_{ADT}= 0 \,.  
\end{align}   

Now, we turn to the boundary formalism.  In this case,  we choose  counter terms of the scalar field as
\beq
\Phi_{ct} = -\frac{1}{4}\phi^2 - \frac{1}{96}  \phi^4 \,.
\eeq
By using this form of counter terms, one can see that
\begin{align}
	\label{}
\sqrt{-\g} (T_{G})^{i}\,_{j} = & \left[
 \frac{1}{8} e^{\eta} \phi_{\textrm{\tiny (0)}}^2   -\frac{3}{128} \phi_{\textrm{\tiny (0)}}^4 +  \frac{3(1+\nu)}{512} \phi_{\textrm{\tiny (0)}}^4 \right] \d^i_j\,- \frac{3(1+\nu)}{256} \phi_{\textrm{\tiny (0)}}^4 \delta^{it}\delta_{jt}\,, \\
	\sqrt{-\g} (T_{\phi})^{i}\,_{j} =       & -\left[
 \frac{1}{8} e^{\eta} \phi_{\textrm{\tiny (0)}}^2  -\frac{3}{128} \phi_{\textrm{\tiny (0)}}^4   \right] \d^i_j\,,                                                                                                                             \\
	\sqrt{-\gamma} \Pi_{\phi} =             & \,0\,.
\end{align}
Once again,		 it is straightforward to confirm the equivalence relation (\ref{equivalence}) for Killing vectors $\xi_T$ and $\xi_R$. As a result, the identical expression for the mass and  angular momentum can be obtained through the boundary stress tensor method as well. Furthermore, one can see  that  each leading divergent term in $Q_{ADT}(\delta g)$ and $ Q_{ADT}(\delta \phi)$ matches with the corresponding one in $\delta(\sqrt{-\gamma}T_G)$ and $\delta(\sqrt{-\gamma}T_{\phi})$, respectively.
It is amusing to note that each ADT potential $Q^{\eta i}_{ADT}(\xi)$ is proportional to the corresponding Killing vector $\xi$, which is not clear a priori from the bulk formalism. This seems  natural from the equivalence relation since the boundary stress tensor $(T_{B})^i_j$ for the static black holes becomes diagonal.

\subsection{General $d$-dimensional static black holes}
In general $d$ dimensions, we focus on   planar  static  black holes with scalar hair in class I. 
The relevant coefficient in the radial expansion of the metric is given by
\begin{equation} \label{}
\g_{ij}^{ (d)} =  \Big[C - \frac{1}{4(d-1)}\phi_{\textrm{\tiny (0)}}^2 \Big]\eta_{ij} +d C~ \delta_{it}\delta_{jt} \,,
\end{equation}
where $C$ is an arbitrary constant.
By using the expression of the quasi-local ADT potential  given in  Eq.~(\ref{MetricCon}), one can see that
\begin{align} \label{}
\sqrt{-g} Q_{ADT}^{\eta t} &(\xi_T\,; \,\d g  ) \big|_{\eta \rightarrow \infty}=  -\frac{d}{4} \phi_{\textrm{\tiny (0)}} \,\d \phi_{\textrm{\tiny (0)}}+\,\frac{d(d-1)}{2} \d C \, ,\\
\sqrt{-g} Q_{ADT}^{\eta t} &(\xi_T\,; \,\d \phi   ) \big|_{\eta \rightarrow \infty} = \frac{d}{4} \phi_{\textrm{\tiny (0)}} \,\d \phi_{\textrm{\tiny (0)}} \,  
\,.
\end{align}
The full expression of counter terms for the metric field in general $d$-dimensions  is not known explicitly even in Einstein gravity. Yet one may still ignore their contributions to the boundary stress tensor except for the boundary  cosmological constant if the  boundary metric is taken flat, $\gamma^{(0)}_{ij} = \eta_{ij}$. With this assumption, 
the boundary stress tensor is given by
\begin{align} \label{}
\sqrt{-\g}(T_{g})^{tj} \xi^{T}_{j} =& ~\frac{d}{8} \phi_{\textrm{\tiny (0)}}^2 -\frac{d(d-1)}{2}\,C   \,, \\
\sqrt{-\g}(T_{\phi})^{tj} \xi^{T}_{j} =& -\frac{d}{8} \phi_{\textrm{\tiny (0)}}^2    \,, \\
\sqrt{-\gamma} \Pi_{\phi} = &0\,.
\end{align}
Once again, we confirm our general results given in Eq.~(\ref{equivalence}).

The total mass of these black holes is obtained as
\begin{equation}
M = M^{g} + M^{\phi} = \frac{d(d-1)}{16\pi G} V_{d-1} C\,.
\end{equation}
where   $V_{d-1}$ denotes the volume of the $(d-1)$-dimensional planar space.
In class II, it is straightforward to apply our formalism to the known analytic solutions for instance, those given in~\cite{Martinez:2004nb}.

\section{Conclusion}

In this paper we have constructed a quasi-local formalism for conserved charges in a general theory of gravity with diffeomorphism symmetry in the presence of arbitrary matter fields. This construction can be regarded as the full- fledged extension of the covariant formalism developed by Abott, Deser and Tekin, which depends on the Euler-Lagrange expressions only.  While the original ADT formulation incorporates the metric fields only at the asymptotic infinity, our construction incorporates the contribution of slow falloff matter fields and can be applied even in the interior region in the sense of quasi-local conserved charges. 

We have shown that our formalism or the full-fledged extension of the ADT formalism at the quasi-local level gives us completely identical results on potentials as those from the covariant phase space approach. In fact the equivalence of potentials in both formalisms  is proven at the off-shell level. Technically, we have adopted a one-parameter path in the solution space in order to obtain finite conserved charges from the off-shell expression.

For the asymptotically (locally) AdS space, we have also introduced identically conserved boundary currents in the same spirit as in the bulk case and obtained the corresponding conserved charges. We have shown that these charges have the same expression  as those from the conventional holographic approach known as the boundary stress tensor method. Furthermore, we have proved that the bulk formalism on conserved charges leads to the same results as the boundary one  by showing that the bulk off-shell ADT potential reduces to the boundary current when we approach the asymptotic infinity. In all, we have shown that our quasi-local formalism can be matched completely with the previously well-known  methods. As a byproduct of these matchings, we have verified in a general theory of gravity that conserved charges by the covariant phase approach should be identical with those by the holographic method. This result can be regarded as the extension of the proof  on  the equivalence of conserved charges  in Einstein gravity  from the covariant phase space formalism and those from the boundary stress tensor method. 

As an application of our formalism, we have considered some examples in order to show some details in our formalism concretely. The necessity of the matter contribution to conserved charges is manifest in these examples. Through the linear analysis, some additional features on matchings between the quasi-local ADT potential and the boundary stress tensor have been explained.

Our matchings among various approaches to conserved charges clarify some equivocal aspects in each formulation on conserved charges. For instance, the consistency of conserved charges with the first law of black hole thermodynamics is not so manifest in the holographic approach while the finiteness of the ADT potential for the asymptotically AdS geometry is not manifest in the ADT formalism. On the other hand, the consistency of conserved charges with the first law of the black hole thermodynamics is usually taken as the property in the covariant phase space and the finiteness of conserved charges is manifest, by construction,  in the holographic approach. All such equivocal aspects disappear since conserved charges are matched through our construction. 

One may note that the second term in Eq.~(\ref{ADTcpt}) plays essential roles to define conserved charge consistent with known results. The analogous term in the boundary formalism  is the second one in  Eq.~(\ref{BdADTcur}), which is not revealed in  literatures on the boundary stress tensor formalism. By presuming that  the conformal anomaly is invariant along the path, we argue that there is no contribution from the second term in Eq.~(\ref{BdADTcur}), which corresponds to the known results. Indeed, there is no contribution from the second term in all the examples we have presented in this paper.  It is amusing to speculate the case in  which the conformal anomaly is not invariant along the path in the solution space. In that case the second term  in Eq.~(\ref{BdADTcur}) would be essential and our expression of holographic conserved charges would be an improvement over the known one. 

We would like to give some comments on the further extension of our formalism. As mentioned in the previous sections, our bulk quasi-local construction can be applied even to the case when a  bulk Lagrangian contains non-manifestly covariant terms  like gravitational Chern-Simons terms.
Though explicit steps are not presented in the presence of non-manifestly covariant terms in the bulk Lagrangian, it would be straightforward to match our final expressions with those in the covariant phase space approach by modifying it to accommodate such terms~\cite{Tachikawa:2006sz,Borowiec:1998st,Bonora:2011gz, Kim:2013cor}. The equivalence with  holographic methods would also hold in the presence of such terms.
Thouogh  the equivalence between  conserved charges from the bulk and  boundary formalisms is shown by adopting FG coordinates, it is expected to hold in other coordinates. It would be interesting to prove this in general.   In this paper we have focused on exact Killing vectors. It would  also be straightforward to extend our construction  to asymptotic Killing vectors by following steps worked out in~\cite{Hyun:2014kfa}. It would be an interesting direction to extend our equivalence between the bulk and boundary constructions  to geometries which are not asymptotically (locally) AdS space.

\vskip 1cm
\centerline{\large \bf Acknowledgments}
\vskip0.5cm

{SH was supported by the National Research Foundation of Korea(NRF) grant funded 
by the Korea government(MOE) with the grant number  2012046278 and the grant number NRF-2013R1A1A2011548.  S.-H.Yi was supported by the National Research Foundation of Korea(NRF) grant funded by the Korea government(MOE) (No.  2012R1A1A2004410).}

%
%

\section*{A. Derivation of the off-shell identity}
\renewcommand{\theequation}{A.\arabic{equation}}
  \setcounter{equation}{0}
In order to obtain the off-shell identity given in Eq.~(\ref{ida}),  let us  note that the  diffeomorphism variation $\delta_{\zeta}\Psi = \Lie_{\zeta}\Psi$  leads to 
\begin{eqnarray}   \label{}
 \delta_{\zeta} (\sqrt{-g}\CL) 
&=& \sqrt{-g}\Big[ -\CE^{\mu\nu}\Lie_{\zeta}g_{\mu\nu} + \CE_{\psi}\Lie_{\zeta}\psi\Big] + \p_{\mu}{\Theta}^{\mu}(\Lie_{\zeta} \Psi)  \nn \\
&=& \sqrt{-g}\Big[ 2\zeta_{\nu} \nabla_{\mu}\CE^{\mu\nu} + \CE_{\psi}\Lie_{\zeta}\psi \Big]+ \p_{\mu}\Big(\Theta^{\mu}(\Lie_{\zeta}\Psi) - 2\sqrt{-g}\CE^{\mu\nu}\zeta_{\nu}\Big)\,, 
\end{eqnarray}
where we have used $\Lie_{\zeta}g_{\mu\nu} = 2 \nabla_{(\mu}\zeta_{\nu)}$ and performed the integration by parts on the first term.  Alternatively, since the diffeomorphism is the symmetry of the given action, the diffeomorphism variation of the Lagrangian  can be written as a total derivative in the form of
\begin{equation} \label{Diffeo}
\delta_{\zeta}(\sqrt{-g}\CL) =   \p_{\mu}\Big(\zeta^{\mu}\sqrt{-g}\CL +\Sigma^{\mu}(\zeta) \Big)\,,
\end{equation}
where $\Sigma^{\mu}$ denotes an additional surface term  which exists for non-manifestly covariant terms like gravitational Chern-Simons terms. By equating the above two forms of diffeomorphism variation, one can see that 
\begin{equation} \label{Sanga}
\sqrt{-g}\Big[ 2\zeta_{\nu} \nabla_{\mu}\CE^{\mu\nu} + \CE_{\psi}\Lie_{\zeta}\psi \Big]= \p_{\mu}\Big(\zeta^{\mu}\sqrt{-g}\CL +\Sigma^{\mu}(\zeta)- \Theta^{\mu}(\Lie_{\zeta}\Psi) + 2\sqrt{-g}\CE^{\mu\nu}\zeta_{\nu}\Big)\,.
\end{equation}
Since the left hand side of Eq.~(\ref{Sanga}) is composed only of $\zeta$ and $\nabla \zeta$ terms for an arbitrary function $\zeta$, one can deduce that  the right hand side  should be taken in the form of
\[   
{\rm r.h.s.}= \sqrt{-g}\nabla_{\mu}\Big(\CY^{\mu\nu}{\zeta}_{\nu} + \CY^{[\mu\nu]\rho}\nabla_{\nu}{\zeta}_{\rho}  \Big)=  \sqrt{-g}\nabla_{\mu}\Big(\CY^{\mu\nu}\zeta_{\nu}- \nabla_{\nu}\CY^{[\mu\nu]\rho}{\zeta}_{\rho}  \Big)\,,
\]
where we have used $\nabla_{\mu}\nabla_{\nu}(\CY^{[\mu\nu]\rho}\zeta_{\rho})=0$.
As a result,  the off-shell identity follows.

\section*{B.  Formulae for the conservation of currents }
\renewcommand{\theequation}{B.\arabic{equation}}
  \setcounter{equation}{0}

In this appendix we show some formulae which are used for the derivation of the conservation of off-shell currents. One may note that the generic double variations of the bulk action can be written as
\begin{equation} \label{}
\delta_{2} \delta_{1}I [\Psi]= \frac{1}{16\pi G}\int d^{D}x\Big[\delta_{2}\Big(\sqrt{-g}\CE_{\Psi}\delta_{1} \Psi\Big) +\p_{\mu}\delta_{2}\Theta^{\mu}(\delta_{1}\Psi) \Big]\,.
\end{equation}
By using the fact that the anti-symmetrization of double variations of the action vanish, $(\delta_{1}\delta_{2} - \delta_{2}\delta_{1})I=0$ and taking one of the variations as a diffeomorphism variation, one can see that
\begin{equation} \label{}
0 = \frac{1}{16\pi G} \int d^{D}x \Big[\delta_{\zeta}\Big(\sqrt{-g}\CE_{\Psi}\delta \Psi\Big) - \delta\Big(\sqrt{-g}\CE_{\Psi}\delta_{\zeta} \Psi\Big) - \p_{\mu}\omega^{\mu}(\delta \Psi, \delta_{\zeta}\Psi)\Big]\,.
\end{equation}
Since $\delta_{\xi}\Psi =0$ and $\omega^{\mu}(\delta \Psi, \delta_{\xi}\Psi)=0$ for a Killing vector $\xi$, it is straightforward to obtain  the following formula 
\begin{equation} \label{}
\delta_{\xi}\Big(\sqrt{-g}\CE_{\Psi}\delta \Psi\Big) = \p_{\mu}\Big(\xi^{\mu}\sqrt{-g}\CE_{\Psi}\delta \Psi\Big)=0\,.
\end{equation}
Combining this formula with Eq.~(\ref{ADTcon1}), one can check the identical conservation of $\CJ^{\mu}_{ADT}$.

By applying the same argument to the on-shell renormalized action given in Eq.~(\ref{ReAction}) , one can obtain
\begin{equation} \label{}
\p_{i}\bigg[\xi^{i}_{B}\sqrt{-\gamma}\Big(T^{kl}_{B}\delta \gamma_{kl} + \Pi_{\psi}\delta\psi\Big)\bigg]=0\,,
\end{equation}
which is used to show the identical conservation of the boundary current $\CJ^{i}_{B}$.

\newpage


\end{document}